\documentclass[amsfonts,11pt]{article}
\usepackage{mathrsfs}
\usepackage{amsmath}
\usepackage{amsfonts}
\usepackage{amssymb, amsmath, cite}
\usepackage{color}
\usepackage{graphicx}

\usepackage{amsmath, amssymb, amsfonts, amsthm}
\date{}

\newcount\hour \newcount\minute
\hour=\time  \divide \hour by 60
\minute=\time
\loop \ifnum \minute > 59 \advance \minute by -60 \repeat
\def\nowtwelve{\ifnum \hour<13 \number\hour:
                      \ifnum \minute<10 0\fi
                      \number\minute
                      \ifnum \hour<12 \ A.M.\else \ P.M.\fi
         \else \advance \hour by -12 \number\hour:
                      \ifnum \minute<10 0\fi
                      \number\minute \ P.M.\fi}
\def\nowtwentyfour{\ifnum \hour<10 0\fi
                \number\hour:
                \ifnum \minute<10 0\fi
                \number\minute}

\title{Integer-Squared Laws for Global Vortices\\ in the Born--Infeld Wave Equations}

\author{Zhifeng Gao\footnote{Email address: gzf@henu.edu.cn}\\Institute of Contemporary Mathematics\\School of Mathematics\\Henan University\\
Kaifeng, Henan 475004, PR China\\\\Sven Bjarke Gudnason\footnote{Email address: gudnason@keio.jp}\\Department of Physics, and \\Research and
Education Center for Natural Sciences \\Keio University\\Hiyoshi 4-1-1, Yokohama, Kanagawa 223-8521, Japan
\\\\Yisong Yang\footnote{Email address: yisongyang@nyu.edu}\\Courant Institute of Mathematical Sciences\\New York University\\New York, New York 10012, U. S. A.}

\newcommand{\bfR}{{\Bbb R}}

\def\ben{\begin{equation}}
\def \een{\end{equation}}

\def\XXint#1#2#3{{\setbox0=\hbox{$#1{#2#3}{\int}$}
 \vcenter{\hbox{$#2#3$}}\kern-.5\wd0}}

\newtheorem{oldtheorem}{Theorem}[section]
\newtheorem{oldassertion}[oldtheorem]{Assertion}
\newtheorem{oldproposition}[oldtheorem]{Proposition}
\newtheorem{oldremark}[oldtheorem]{Remark}
\newtheorem{oldlemma}[oldtheorem]{Lemma}
\newtheorem{olddefinition}[oldtheorem]{Definition}
\newtheorem{oldclaim}[oldtheorem]{Claim}
\newtheorem{oldcorollary}[oldtheorem]{Corollary}

\newcommand{\dd}{\mbox{d}}
\newcommand{\ee}{\end{equation}}
\newcommand{\be}{\begin{equation}}\newcommand{\bea}{\begin{eqnarray}}
\newcommand{\eea}{\end{eqnarray}}
\newcommand{\ii}{\mbox{i}}\newcommand{\e}{\mbox{e}}
\newcommand{\pa}{\partial}

\newcommand{\nn}{\nonumber}

\newcommand{\lm}{\lambda}

\begin{document}
\maketitle


\begin{abstract}
A series of
quantization identities are established for static vortex solutions governed by the Born--Infeld type actions. These identities are of a universal nature which 
are indifferent to the details of the models and provide refined descriptions of divergent energetic quantities.
\smallskip

\smallskip


\smallskip


\end{abstract}

\maketitle


\section{Introduction}

A well-known puzzle in theoretical physics is the structure of the electron viewed as a classical particle, which, when regarded as a point
charge,  carries infinite energy
in the context of the Maxwell theory of electromagnetism. In order to accommodate the idea of a point charge for the electron, Born and Infeld,
in a series of pioneering papers \cite{Born1,Born2,Born-Infeld1,Born-Infeld2}, developed an elegant nonlinear field theory of electromagnetism, later
called the Born--Infeld theory, and also the Dirac--Born--Infeld theory, which enables a point charge to carry a finite energy. In modern physics, especially in the contemporary research of
string and membrane theories, the Born--Infeld formalism arises frequently in numerous fundamental and profound ways. For example, it naturally originates from
the Nambu--Goto string theory; via the Seiberg--Witten map, its
action may be seen to take place in the formalism of
 non-commutative field theories \cite{BY,Gar}; in the context of quantum gravity theory aimed to resolve the celebrated singularity and non-unitarity problems plaguing general
relativity, it gives rise to deep insight and new ideas as to how spacetime geometry and matter-interaction may be coupled; and, with
 supersymmetric extensions, its action appears in brane theories and supergravity study.
See \cite{EMY} for a recent short historical account of the Born--Infeld theory. Thus, the Born--Infeld theory has become a basic construct of quantum field
theory and its modern extensions and modifications. Mathematically, however, the structure of the Born--Infeld action is difficult and challenging, so that the progress
towards its full understanding has been slow and staggering, although as in the Abelian Higgs model \cite{JT} some precise and exact existence results are available
 \cite{Y1,Y2}
in the BPS \cite{Bo,PS} critical phase \cite{SH} (see \cite{CH,Han} for some recent extended investigations along this line). The main result of this work is to establish an interesting
family of topological laws associated with global vortex solutions of various Born--Infeld models which may be expressed universally in terms of the square of the vortex charge.

Global vortices are topological defects in two-spatial dimensions, which may also be viewed as global strings in three-spatial dimensions, whose strikingly simple
topological characterization owes itself to the spontaneously broken $U(1)$ symmetry of the vacuum manifold of the theory. Nonrelativistic global vortices 
are known to arise
in superfluids and governed by the Gross--Pitaevskii equation \cite{Gross,Pit}, and relativistic global vortices are known to arise as the mixed states or the Abrikosov
vortices \cite{Ab} in
type-II superconductors and governed by the Ginzburg--Landau equation \cite{GL}. In theoretical cosmology, global vortices, as well as local vortices, may be
created in the early universe as topological remnants resulting from a series of phase transitions, due to expansion and cooling of the universe, to appear as
 spots of concentrated energy and curvature distribution, around which matter accretion may take place, thus providing a mechanism for galaxy formation
\cite{GS,K1,K2,Mar,V,VS}. Not surprisingly, mathematical studies on the global vortices are dominately centered around their topological characterization realized
as the winding number, say $N$, of the phase of the order parameter, also represented as the total vortex number. Due to the absence of a gauge field
responsible for restoring local symmetry, the kinetic part of the vortex energy diverges logarithmically but its Higgs potential part remains finite and is
elegantly and dramatically
given by a quantity which is proportional to $N^2$. Such a property is signaturally known in the trade as the quantization
\cite{BMR1,BMR2} of the Ginzburg--Landau energy which has propelled enormous subsequent studies \cite{BBH,Br,Lin1,Lin2,PT,SS}.
 In the present work we unveil the same quantization property for global vortices in a series of actively investigated Born--Infeld models.

Interestingly, the structure of the energy-momentum tensor of the Born--Infeld action renders a peculiar static virial theorem in the form $E=E_0$ where
$E$ is the total energy of the theory 
 and $E_0$ is a certain `derived energy' resulting from the conformal property
of the kinetic and elastic part of the action. However, for an $N$-vortex solution, both $E$ and $E_0$ are actually divergent and the virial theorem is trivially valid.
The main theme of  this work is to show that in the sense of taking a full-plane limit there holds the universal identity of the form
\be\label{1.1}
E-E_0=\frac{\pi N^2}2,\quad N=0,1,2,\dots.
\ee
By `universal' we mean that the identity is valid regardless of the fine details of the action of the theory but expressed explicitly, up to the factor of $\frac{\pi}2$, 
or such,
in an integer-squared form.

An outline of the rest of the paper is as follows. In Section 2, we start from a discussion of the wave equation of the Born--Infeld dynamics and the associated virial 
theorem in general dimensions and the quantization property of the Ginzburg--Landau equation. We then derive our quantization identity for the Born--Infeld
wave equation of the form (\ref{1.1}). In Section 3, we extend our study on generalized Born--Infeld models and give some concrete examples which are
beyond the reach of the result of the previous section. In Section 4, we establish the same quantization phenomenon for $N$-vortex solutions arising in
various Born--Infeld models studied in cosmology. In Section 5, we develop our quantization identity for vortices in the Born--Infeld tachyon models. In Section 6,
we present a series of numerical studies to illustrate our theoretical results. In Section 7, we summarize our work in general terms.

\section{The Born--Infeld wave equations and vortices}
\setcounter{equation}{0}

We first consider the Born--Infeld waves in the standard Minkowski spacetime $\bfR^{1,n}$ equipped with the Lorentz metric $(\eta_{\mu\nu})=\mbox{diag}\{1,-1,\dots,-1\}$. The Lagrangian action density now assumes the form
\be\label{2.1}
{\cal L}=b^2\left(1-\sqrt{1-\frac1{b^2}\pa_\mu\phi\pa^\mu\overline{\phi}}\,\right)-V(|\phi|^2),
\ee
where $\phi$ is a complex scalar field,  $V\geq0$ a potential density function, and $b>0$ a constant called the Born coupling parameter. The Euler--Lagrange
equation associated with (\ref{2.1}) is
\be\label{eq1}
\frac12\,\pa_\mu\left(\frac{\pa^\mu\phi}{\sqrt{1-\frac1{b^2}\pa_\nu\phi\pa^\nu\overline{\phi}}}\right)+V'(|\phi|^2)\phi=0,
\ee
along with the induced energy-momentum tensor (also known as the Belinfante--Rosenfeld symmetrized energy-momentum tensor)
\be
T_{\mu\nu}=\frac{\left(\pa_\mu\phi\pa_\nu\overline{\phi}+\pa_\mu\overline{\phi}\pa_\nu\phi
\right)}{2\sqrt{1-\frac1{b^2}\pa_\gamma\phi\pa^\gamma\overline{\phi}}}-\eta_{\mu\nu}{\cal L},
\ee
which satisfies the conservation law $\pa^\nu T_{\mu\nu}=0$.
In this work we are interested in static waves. So (\ref{eq1}) becomes
\be\label{2.3}
\frac12\,\nabla\cdot\left(\frac{\nabla\phi}{\sqrt{1+\frac1{b^2}|\nabla\phi|^2}}\right)=V'(|\phi|^2)\phi,\quad x\in\bfR^n.
\ee
which is the Euler--Lagrange equation of the total static energy given by
\be\label{2.2}
E(\phi)=\int_{\bfR^n}T_{00}\,\dd x=\int_{\bfR^n}\left\{b^2\left(\sqrt{1+\frac1{b^2}|\nabla\phi|^2}-1\right)+V(|\phi|^2)\right\}\,\dd^n x.
\ee
Since the static conservation law assumes the form $\pa_j T_{ij}=0$, an integration of it leads to the partition identity
$\int_{\bfR^n} T_{ii}\,\dd x=0$ (cf. \cite{JT}), or
\be\label{2.5}
\int_{\bfR^n}\frac{|\nabla\phi|^2}{\sqrt{1+\frac1{b^2}|\nabla\phi|^2}}\,\dd x=n\int_{\bfR^n}
\left\{b^2\left(\sqrt{1+\frac1{b^2}|\nabla\phi|^2}-1\right)+V(|\phi|^2)\right\}\,\dd^n x,
\ee
for a finite-energy solution of (\ref{2.3}). Alternatively, (\ref{2.5}) may also be derived by the well-known rescaling argument of Derrick \cite{D} with setting
\be
\left(\frac{\dd}{\dd\sigma} E(\phi_\sigma)\right)_{\sigma=1}=0,
\ee
where $\phi_\sigma(x)=\phi(\sigma x)$ ($\sigma\in\bfR$) and $\phi$ is a finite-energy solution of (\ref{2.3}) or critical point of (\ref{2.2}).
Thus, unlike in classical Klein--Gordon wave situation where the Derrick theorem excludes the spatial dimensions $n\neq1$
(when $n=2$, the potential density needs to be trivial, $V=0$), we see that no dimension is excluded in the Born--Infeld wave formalism, at least at
the Derrick energy-partition level. We now explain and clarify the validity of (\ref{2.5}) for vortex solutions in two dimensions.

To proceed, we first recall the important case where $V$ is of the Ginzburg--Landau type:
\be\label{2.6}
V(|\phi|^2)=\frac\lm8(|\phi|^2-1)^2,
\ee
where $\lm>0$ is a constant. In the Klein--Gordon model case, the Lagrangian density is
\be\label{2.7}
{\cal L}=\frac12\pa_\mu\phi\pa^\mu\overline{\phi}-V(|\phi|^2),
\ee
and the static equation of motion reads
\be\label{2.8}
\Delta \phi=\frac\lm2(|\phi|^2-1)\phi,
\ee
which is also known as the Ginzburg--Landau equation in the absence of a magnetic field, also called the {\em bare} Ginzburg--Landau equation. By the Derrick theorem, (\ref{2.8}) has no nontrivial finite-energy solution over
$\bfR^n$ for any $n\geq2$. However,  it is also known over $\bfR^2$ that the solutions of (\ref{2.8}) enjoy
the potential energy quantization property \cite{BMR1,BMR2}:
\be\label{GL}
\int_{\bfR^2}(|\phi|^2-1)^2\,\dd^2 x=\frac{4\pi N^2}\lm,\quad N=0,1,2,\dots,\infty,
\ee
 (cf. \cite{GGY} for some extended versions of energy conservation laws along the above fashion) which indicates that the energy blow-up of a nontrivial solution of (\ref{2.8}) of finite potential energy must occur in its kinetic energy part:
\be
\int_{\bfR^2}\frac12|\nabla\phi|^2\,\dd^2 x=\infty.
\ee
In fact, a prototype solution of the above characteristics is given by the soliton  solution of an $N$-vortex type 
of  the radially symmetric ``spiral" form \cite{H}
\be\label{2.11}
\phi(x)=u(r)\e^{\ii N\theta},
\ee
where $r,\theta$ are the polar coordinates of $\bfR^2$, $N$ is an integer, and $u$ a real-valued amplitude function, or profile function, satisfying the boundary condition
\be\label{2.14}
u(0)=0,\quad u(\infty)=1.
\ee
Standard analysis also establishes the following precise properties
\be
u(r)=\mbox{O}( r^N)\quad\mbox{for $r$ small}; \quad u(r)=1+\mbox{O}(\e^{-\sqrt{\lm}r})\quad\mbox{for $r$ large}.
\ee
Inserting (\ref{2.6}) and (\ref{2.11}) into (\ref{2.2}) (with $n=2$), we arrive at the reduced energy
\be\label{2.12}
E(u)=\pi\int_0^\infty\left\{2b^2\left(\sqrt{1+\frac1{b^2}\left[u_r^2+\frac{N^2}{r^2}u^2\right]}-1\right)+\frac\lm4(u^2-1)^2\right\}\,r\dd r,
\ee
where and in the sequel, we use $u_r$ to denote the derivative of $u$ with respect to $r$.
Besides, in the present $N$-vortex solution context, (\ref{2.5})
becomes
\bea\label{2.15}
&&\int_0^\infty\frac{\left(u_r^2+\frac{N^2}{r^2}u^2\right)}{\sqrt{1+\frac1{b^2}\left[u_r^2+\frac{N^2}{r^2}u^2\right]}}\,r\,\dd r\nn\\
&&=\int_0^\infty\left\{ 2b^2\left(\sqrt{1+\frac1{b^2}\left[u_r^2+\frac{N^2}{r^2}u^2\right]}-1\right)+\frac\lm4(u^2-1)^2\right\}\,r\dd r,
\eea
so that the Born--Infeld wave equation (\ref{2.3}) reads
\be\label{2.13}
\frac{\dd}{\dd r}\left(\frac{r u_r}{\sqrt{1+\frac1{b^2}\left[u_r^2+\frac{N^2}{r^2}u^2\right]}}\right)=
\frac{N^2 u}{r\sqrt{1+\frac1{b^2}\left[u_r^2+\frac{N^2}{r^2}u^2\right]}}+\frac\lm2\,r(u^2-1)u,\quad r>0,
\ee
which is also the Euler--Lagrange equation of the energy (\ref{2.12}).

However, since $u(\infty)=1$, both sides of (\ref{2.15}) diverge. Thus (\ref{2.15}) is only a formal, ``infinity equals infinity",  relation. Below we elaborate
 on the two sides of (\ref{2.15})
and establish a 
hidden quantization relation which roughly says that the difference of the left-hand and right-hand sides of (\ref{2.15}) is
exactly the half of the square of the vortex number. More precisely, we have
\bea\label{2.24a}
&&\lim_{R\to\infty}\int_0^R\left\{2b^2 \left(\sqrt{1+\frac1{b^2}\left[u_r^2+\frac{N^2}{r^2}u^2\right]}-1\right)+\frac\lm4(u^2-1)^2 -
\frac{\left(u_r^2+\frac{N^2}{r^2}u^2\right)}{\sqrt{1+\frac1{b^2}\left[u_r^2+\frac{N^2}{r^2}u^2\right]}}\right\}\,r\dd r\nn\\
&&=\frac{N^2}2.
\eea

We now proceed to prove (\ref{2.24a}). Indeed, multiplying (\ref{2.13}) with $ru_r$, we have 
\be\label{2.16}
ru_r \left(\frac{r u_r}{\sqrt{1+\frac1{b^2}\left[u_r^2+\frac{N^2}{r^2}u^2\right]}}\right)_r=
\frac{N^2 u u_r}{\sqrt{1+\frac1{b^2}\left[u_r^2+\frac{N^2}{r^2}u^2\right]}}+\frac\lm2\,r^2(u^2-1)uu_r,\quad r>0.
\ee

We first note that the left-hand side of (\ref{2.16}) reads
\bea\label{2.17}
ru_r \left(\frac{r u_r}{\sqrt{1+\frac1{b^2}\left[u_r^2+\frac{N^2}{r^2}u^2\right]}}\right)_r&=&
\left(\frac{r^2u_r^2}{\sqrt{1+\frac1{b^2}\left[u_r^2+\frac{N^2}{r^2}u^2\right]}}\right)_r\nn\\
&&-\left(\frac{r u_r}{\sqrt{1+\frac1{b^2}\left[u_r^2+\frac{N^2}{r^2}u^2\right]}}\right)(u_r+ru_{rr}).
\eea
Furthermore we have
\bea\label{2.18}
b^2\left(r^2\left[ \sqrt{1+\frac1{b^2}\left[u_r^2+\frac{N^2}{r^2}u^2\right]}-1\right]\right)_r&=&\frac{r^2}{\sqrt{1+\frac1{b^2}\left[u_r^2+\frac{N^2}{r^2}u^2\right]}}
\left(u_r u_{rr}+\frac{N^2}{r^2} uu_r-\frac{N^2}{r^3} u^2\right)\nn\\
&&+2b^2 r\left(\sqrt{1+\frac1{b^2}\left[u_r^2+\frac{N^2}{r^2}u^2\right]}-1\right).
\eea
Combining (\ref{2.16})--(\ref{2.18}) we arrive at
\bea
&&\left(\frac{r^2u_r^2}{\sqrt{1+\frac1{b^2}\left[u_r^2+\frac{N^2}{r^2}u^2\right]}}\right)_r
-\left(\frac{r u_r}{\sqrt{1+\frac1{b^2}\left[u_r^2+\frac{N^2}{r^2}u^2\right]}}\right)(u_r+ru_{rr})\nn\\
&=&b^2\left(r^2\left[ \sqrt{1+\frac1{b^2}\left[u_r^2+\frac{N^2}{r^2}u^2\right]}-1\right]\right)_r-
\frac{r^2}{\sqrt{1+\frac1{b^2}\left[u_r^2+\frac{N^2}{r^2}u^2\right]}}
\left(u_r u_{rr}-\frac{N^2}{r^3} u^2\right)\nn\\
&&-2b^2 r\left(\sqrt{1+\frac1{b^2}\left[u_r^2+\frac{N^2}{r^2}u^2\right]}-1\right)+\frac\lm2\,r^2(u^2-1)uu_r,
\eea
which may be simplified to give us the relation
\bea\label{2.20}
&&\left(\frac{r^2u_r^2}{\sqrt{1+\frac1{b^2}\left[u_r^2+\frac{N^2}{r^2}u^2\right]}}\right)_r
=b^2\left(r^2 \left[\sqrt{1+\frac1{b^2}\left[u_r^2+\frac{N^2}{r^2}u^2\right]}-1\right]\right)_r\nn\\
&&+\frac{r\left(u_r^2+\frac{N^2}{r^2}u^2\right)}{\sqrt{1+\frac1{b^2}\left[u_r^2+\frac{N^2}{r^2}u^2\right]}}
-2b^2 r\left(\sqrt{1+\frac1{b^2}\left[u_r^2+\frac{N^2}{r^2}u^2\right]}-1\right)+\frac\lm2\,r^2(u^2-1)uu_r.\nn\\
&&
\eea

We now study the integration of (\ref{2.20}) over the interval $(0,R)$ for $R>0$ large.

First we have
\be\label{2.21}
\int_0^R\left(\frac{r^2u_r^2}{\sqrt{1+\frac1{b^2}\left[u_r^2+\frac{N^2}{r^2}u^2\right]}}\right)_r\,\dd r=\frac{R^2u_r^2(R)}{\sqrt{1+\frac1{b^2}\left[u_r^2(R)+\frac{N^2}{R^2}u^2(R)\right]}},
\ee
which tends to zero as $R\to\infty$ since $u(r)\to1$ sufficiently fast as $r\to\infty$.

Next we have
\bea\label{2.22}
&&\int_0^R b^2\left(r^2 \left[\sqrt{1+\frac1{b^2}\left[u_r^2+\frac{N^2}{r^2}u^2\right]}-1\right]\right)_r\,\dd r\nn\\
&=&b^2\left(R^2 \left[\sqrt{1+\frac1{b^2}\left[u_r^2(R)+\frac{N^2}{R^2}u^2(R)\right]}-1\right]\right)\nn\\
&=&\frac{N^2}2+\mbox{O}(R^{-2}),
\eea
since $u(r)\to1$ as $r\to\infty$ sufficiently fast.

Lastly we have
\be\label{2.23}
\int_0^R r^2 (u^2-1)uu_r\,\dd r=\frac{R^2}4(u^2(R)-1)^2-\frac12\int_0^R (u^2-1)^2 r\,\dd r.
\ee

Therefore, integrating (\ref{2.20}) over $(0,R)$ and applying (\ref{2.21})--(\ref{2.23}), we obtain
\bea\label{2.24}
&&\int_0^R\left\{2b^2 \left(\sqrt{1+\frac1{b^2}\left[u_r^2+\frac{N^2}{r^2}u^2\right]}-1\right)+\frac\lm4(u^2-1)^2 -
\frac{\left(u_r^2+\frac{N^2}{r^2}u^2\right)}{\sqrt{1+\frac1{b^2}\left[u_r^2+\frac{N^2}{r^2}u^2\right]}}\right\}\,r\dd r\nn\\
&=&\frac{N^2}2+\mbox{O}(R^{-2}),
\eea
for $R>0$ sufficiently large, which has already been observed numerically earlier.

It is worth noting that our quantization identity (\ref{2.24a}) or (\ref{2.24}) is independent of the specific form of the potential function (\ref{2.6}). Such
a point will further be illustrated in the next section.

\section{Generalized models and integral identities}
\setcounter{equation}{0}

We now consider the generalized Born--Infeld wave equation, governing a complex scalar field, $\phi$, associated with the Lagrangian density
\be\label{3.1}
{\cal L}=F(X)-V(|\phi|^2),
\ee
as studied in numerous works including \cite{Adam,AB,AL,Ba1,Ba2,BH,BLL,BLMO,RS}, where $X=\frac12\pa_\mu\phi\pa^\mu\overline{\phi}$, which includes as its special examples the standard wave equation
model, the Born--Infeld wave equation model, and the logarithmic extension  \cite{FFP} of the Born--Infeld radical-root formalism
where 
\be\label{3.2}
F(X)={\Lambda^2}\ln\left(1+\frac X{\Lambda^2}\right), 
\ee
$\Lambda^2>0$ being a cutoff energy, among many others. The wave equation or the Euler--Lagrange equation associated with (\ref{3.1}) reads
\be\label{3.3}
\frac12\pa_\mu\left(F_X\pa^\mu \phi\right)+\frac{\pa V}{\pa \overline{\phi}}=0,
\ee
with $F_X=F'(X)$. Note also that the (symmetrized) energy-momentum tensor of (\ref{3.1}) assumes the form
\be\label{3.4}
T_{\mu\nu}=\frac12\left(\pa_\mu\phi\pa_\nu\overline{\phi}+\pa_\mu\overline{\phi}\pa_\nu\phi\right)F_X-\eta_{\mu\nu}{\cal L}.
\ee
So, in the static limit, we obtain the energy density
\be\label{3.5}
{\cal E}=T_{00}=-F\left(-\frac12|\nabla\phi|^2\right)+V(|\phi|^2),
\ee
the reduced equation of motion
\be\label{3.6}
\frac12\nabla\cdot\left(F_X \nabla\phi\right)=V'(|\phi|^2)\phi,
\ee
and the energy-partition relation
\be
\int_{\bfR^n} F_X\!\left(-\frac12|\nabla\phi|^2\right) |\nabla\phi|^2\,\dd^n x=n\int_{\bfR^n}\left\{-F\!\left(-\frac12|\nabla\phi|^2\right)+V(|\phi|^2)\right\}\,\dd^n x.
\ee

We next specialize on two-spatial dimensions and specify the $N$-vortex ansatz (\ref{2.11}). Hence we have
\be
X=-\frac12\left(u_r^2+\frac{N^2}{r^2} u^2\right),
\ee
so that we get from (\ref{3.6}) the generalized Born--Infeld vortex equation
\be\label{3.8}
\left(r F_X u_r\right)_r-F_X\frac{N^2}r u-2r V'(u^2) u=0,
\ee
and obtain as well the associated formal virial formula
\bea\label{310}
&&\int_0^\infty F_X\!\left(-\frac12\left[u_r^2+\frac{N^2}{r^2}u^2\right]\right)\! \left(u_r^2+\frac{N^2}{r^2}u^2\right)\,r\dd r\nn\\
&&=2\int_0^\infty
\left\{-F\!\left(-\frac12 \left[u_r^2+\frac{N^2}{r^2}u^2\right]\right)+V(u^2)\right\}\,r\dd r.
\eea
Note that, again, such a relation is only formally valid, since it universally indicates an equality between two infinite quantities, whose precise content
will be elaborated and illustrated as follows.

First, we obtain from (\ref{3.8}) the relation
\be\label{3.9}
ru_r\left(F_X ru_r\right)_r=F_X N^2 uu_r+2r^2 V'(u^2) uu_r.
\ee
Moreover, we can rewrite the left-hand side of (\ref{3.9}) as
\bea\label{3.10}
ru_r(F_X ru_r)_r&=&(F_X r^2 u_r^2)_r-F_X ru_r (ru_r)_r\nn\\
&=&(F_X r^2 u_r^2)_r-F_X(ru^2_r+r^2 u_r u_{rr}).
\eea
Besides, there holds
\be\label{3.11}
(r^2 F)_r=-\left(N^2 uu_r+r^2 u_r u_{rr}-\frac{N^2}r u^2\right)F_X+2rF.
\ee
Hence, in view of  (\ref{3.11}), we may express the right-hand side of (\ref{3.9}) as
\be\label{3.12}
F_X N^2 uu_r+2r^2 V'(u^2) uu_r=-(r^2 F)_r-F_X r^2 u_r u_{rr}+\frac{N^2}r F_X u^2+2r F+2r^2 V'(u^2)uu_r.
\ee
Combining (\ref{3.10}) and (\ref{3.12}), we arrive at
\be\label{3.13}
(F_X r^2 u_r^2)_r=-(r^2 F)_r  +\left(ru^2_r+\frac{N^2}r u^2\right)F_X+2r F+2r^2 V'(u^2)uu_r.
\ee

We then integrate (\ref{3.13}) over $(0,R)$ where $R>0$ is sufficiently large term by term and study each term involved. First, we have
\be\label{3.14}
\int_0^R r^2 V'(u^2) u u_r\,\dd r=\frac12 R^2 V(u^2(R))-\int_0^R V(u^2(r))\,r\dd r.
\ee
Naturally, we may assume that $u(r)$ approaches its vacuum value $u_\infty>0$ sufficiently fast so that
\be\label{3.15}
u(r)-u_\infty=\mbox{O}(r^{-2})\quad\mbox{for $r>0$ large}.
\ee

Moreover, since $V$ is differentiable and resembles a Mexican-hat type potential density, we may assume 
\be\label{3.16}
V(u_\infty^2)=0, \quad V(u^2)=\mbox{O}((u-u_\infty)^2)\quad \mbox{for $u$ near $u_\infty$}.
\ee
Thus by (\ref{3.15}) and (\ref{3.16}) we have $V(u(R))=\mbox{O}(R^{-4})$ for $R>0$ large. Inserting this into (\ref{3.14}), we get
\be\label{3.17}
\lim_{R\to\infty}\int_0^R r^2 V'(u^2) uu_r\,\dd r=-\int_0^\infty V(u^2(r))\,r\dd r.
\ee

Furthermore, we have
\be\label{3.18}
\int_0^R (F_X r^2 u_r^2)_r\,\dd r=F_X\left(-\frac12\left[u_r^2(R)+\frac{N^2}{R^2} u^2(R)\right]\right) R^2 u_r^2 (R)\to 0\quad\mbox{as }R\to\infty.
\ee
Besides, in order to recover the classical wave scalar model (\ref{2.7}) in the weak field limit, $X\sim 0$, we assume
\be\label{3.19}
F(X)=X+\mbox{O}(X^2)\quad\mbox{for $X$ near zero}.
\ee
Thus, from (\ref{3.19}), we find
\be\label{3.20}
\int_0^R (r^2 F)_r\,\dd r=-\frac{N^2 u_\infty^2}2+\mbox{O}(R^{-1})\quad\mbox{as }R\to\infty.
\ee
Integrating (\ref{3.13}) over $(0,R)$ and letting $R\to\infty$, along with the results (\ref{3.17}), (\ref{3.18}), and (\ref{3.20}), we obtain
\be\label{3.21}
\lim_{R\to\infty}\int_0^R\left(-2F+2V(u^2)-\left[u^2_r+\frac{N^2}{r^2} u^2\right]F_X\right)\,r\dd r=\frac{N^2 u_\infty^2}2,
\ee
which clarifies the meaning of the equality (\ref{310}) and includes (\ref{2.24}) as a classical special case with
\be
F(X)=b^2\left(1-\sqrt{1-\frac2{b^2}X}\right).
\ee
Moreover, when $V$ is given by (\ref{2.6}) so that $u_\infty=1$ and $F$ is given by (\ref{3.2}), we have
\bea\label{3.23}
&&\lim_{R\to\infty}\int_0^R\left\{\frac\lm 4(u^2-1)^2-
2\Lambda^2\ln\left(1-\frac1{2\Lambda^2}\left[u_r^2+\frac{N^2}{r^2}u^2\right]\right)
-\frac{\Lambda^2\left(u_r^2+\frac{N^2}{r^2}u^2\right)}{\Lambda^2-\frac1{2}\left(u_r^2+\frac{N^2}{r^2}u^2\right)}\right\}\,r\dd r\nn\\
&&=\frac{N^2}2,
\eea
where $u$ is an $N$-vortex solution governed by the equation (\ref{3.8}) in its correspondingly reduced form, namely,
\be\label{326}
\left(\frac{\Lambda^2}{\Lambda^2-\frac12\left(u_r^2+\frac{N^2}{r^2}u^2\right)} ru_r\right)_r=\frac{\Lambda^2}{\Lambda^2-\frac12\left(u_r^2+\frac{N^2}{r^2}u^2\right)}\,\frac{N^2} r u
+\frac\lm2 r(u^2-1)u.
\ee
Furthermore, for the following power-law type kinetic energy density
\be\label{3.24}
F(X)=X+X^3,
\ee
taken in \cite{Ba2}, with $V$ given in (\ref{2.6}), we see that the vortex equation (\ref{3.8}) assumes the form
\be\label{328}
\left(r\left(1+\frac34\left[u_r^2+\frac{N^2}{r^2}u^2\right]^2\right)u_r\right)_r=\left(1+\frac34\left[u_r^2+\frac{N^2}{r^2}u^2\right]^2\right)
\frac{N^2}r u+\frac\lm2 r(u^2-1)u,
\ee
and we  see that the identity (\ref{3.21}) becomes
\bea\label{3.25}
&&\lim_{R\to\infty}\int_0^R\left\{u_r^2+\frac{N^2}{r^2}u^2 +\frac14\left( u_r^2+\frac{N^2}{r^2}u^2 \right)^3+\frac\lm4(u^2-1)^2\right.\nn\\
&&\left.\quad\quad-\left(1+\frac34\left[u_r^2+\frac{N^2}{r^2}u^2 \right]^2\right)\left(u_r^2+\frac{N^2}{r^2}u^2\right)\right\}\,r\dd r\nn\\
&&=\lim_{R\to\infty}\int_0^R\left\{\frac\lm4(u^2-1)^2
-\frac12\left( u_r^2+\frac{N^2}{r^2}u^2 \right)^3\right\}\,r\dd r =\frac{N^2}2.
\eea

We note that the left-hand sides of (\ref{3.23}) and (\ref{3.25}) are again expressed in terms of the differences of two positively divergent quantities, as in (\ref{2.24a}) or (\ref{2.24}).

\section{Dirac--Born--Infeld vortices arising in cosmology}
\setcounter{equation}{0}

Following \cite{AST,ST}, we consider the Dirac--Born--Infeld action density arising in inflationary cosmology and string theory given by
\be\label{4.1}
{\cal L}=\frac1{f(|\phi|^2)}\left(1-\sqrt{1-2f(|\phi|^2)X}\right)-V(|\phi|^2),
\ee
governing a complex scalar field $\phi$, again with $X=\frac12\pa_\mu\phi\pa^\mu\overline{\phi}$, where $f$ is a positive-valued function.
See also \cite{AKL,Copeland,GO,MS} and references therein. The equation of motion of (\ref{4.1}) is
\bea\label{4.2}
\frac12\,\pa_\mu\left(\frac{\pa^\mu\phi}{\sqrt{1-2f(|\phi|^2)X}}\right)&=&\frac{f'(|\phi|^2)X\phi}{f(|\phi|^2)\sqrt{1-2f(|\phi|^2)X}}\nn\\
&&-\frac{f'(|\phi|^2)\phi}{f^2(|\phi|^2)}\left(1-\sqrt{1-2f(|\phi|^2)X}\right)-V'(|\phi|^2)\phi.
\eea
The associated (symmetrized) energy-momentum tensor is
\be\label{4.3}
T_{\mu\nu}=\frac{\pa_\mu\phi\pa_\nu\overline{\phi}+\pa_\mu\overline{\phi}\pa_\nu\phi}{2\sqrt{1-2f(|\phi|^2)X}}-\eta_{\mu\nu}{\cal L}.
\ee
Thus, in static situation, the energy density assumes the form
\be\label{4.4}
{\cal E}=\frac1{f(|\phi|^2)}\left(\sqrt{1+f(|\phi|^2)|\nabla\phi|^2}-1\right)+V(|\phi|^2),
\ee
and the equation of motion (\ref{4.2}) becomes
\bea\label{4.5}
\nabla\cdot\left(\frac{\nabla\phi}{\sqrt{1+f(|\phi|^2)|\nabla\phi|^2}}\right)&=&\frac{f'(|\phi|^2)|\nabla\phi|^2\phi}{f(|\phi|^2)\sqrt{1+f(|\phi|^2)|\nabla\phi|^2}}\nn\\
&&-\frac{2f'(|\phi|^2)\phi}{f^2(|\phi|^2)}\left(\sqrt{1+f(|\phi|^2)|\nabla\phi|^2}-1\right)+2V'(|\phi|^2)\phi.
\eea
As before, we may also obtain the virial relation
\be\label{4.6}
\int_{\bfR^n}\frac{|\nabla\phi|^2}{\sqrt{1+f(|\phi|^2)|\nabla\phi|^2}}\,\dd^n x =n\int_{\bfR^n} \left(\frac1{f(|\phi|^2)}\left(\sqrt{1+f(|\phi|^2)|\nabla\phi|^2}-1\right)+V(|\phi|^2)\right)\,\dd^n x.
\ee

After the above general discussion, we now focus on global vortex solutions with $n=2$ again, specified within the framework of the ansatz (\ref{2.11}), subject to
the asymptotic behavior (\ref{3.15}) dictated by a spontaneously broken symmetry. For convenience, we use the compact notation
\be
Y=|\nabla \phi|^2=u_r^2+\frac{N^2}{r^2}u^2.
\ee
Then the equation of motion (\ref{4.5}) becomes
\be\label{4.8}
\left(\frac{ru_r}{\sqrt{1+fY}}\right)_r=\frac{N^2 u}{r\sqrt{1+fY}}-\frac{2r f' u}{f^2}\left(\sqrt{1+fY}-1\right)+\frac{rf'u Y}{f\sqrt{1+fY}}+2rV' u,
\ee
and the virial relation (\ref{4.6}) is specialized into
\be\label{4.9}
\int_0^\infty \frac{rY}{\sqrt{1+fY}}\,\dd r=2\int_0^\infty\left(\frac1f\left(\sqrt{1+fY}-1\right)+V\right) r\dd r.
\ee
Since the properties of the vortex profile function $u$ implies the asymptotic estimate
\be\label{4.10}
Y(r)=\frac{N^2 u^2_\infty}{r^2}+\mbox{O}(r^{-3}),\quad r\to\infty,
\ee
we see that both sides of (\ref{4.9}) diverges. Therefore, as before, we need more elaboration on such a relation.

First, multiplying both sides of (\ref{4.8}) by $ru_r$, we get
\be\label{4.11}
ru_r\left(\frac{ru_r}{\sqrt{1+fY}}\right)_r=\frac{N^2 u u_r}{\sqrt{1+fY}}-\frac{2r^2 f' uu_r}{f^2}\left(\sqrt{1+fY}-1\right)+\frac{r^2f' Yuu_r}{f\sqrt{1+fY}}+2r^2V' uu_r.
\ee
Next, we note that
\bea\label{4.12}
\left(\frac{r^2}f\left(\sqrt{1+fY}-1\right)\right)_r&=&\frac{r^2 u_r u_{rr}}{\sqrt{1+fY}}+\frac{N^2 uu_r}{\sqrt{1+fY}}-\frac{N^2 u^2}{r\sqrt{1+fY}}\nn\\
&&+\frac{r^2 f' Y uu_r}{f\sqrt{1+fY}}+\frac{2r}{f}\left(1-\frac{rf' uu_r}f\right)\left(\sqrt{1+fY}-1\right),
\eea
and that
\be\label{4.13}
ru_r\left(\frac{ru_r}{\sqrt{1+fY}}\right)_r=\left(\frac{r^2u_r^2}{\sqrt{1+fY}}\right)_r-\frac{(ru_r^2+r^2 u_r u_{rr})}{\sqrt{1+fY}}.
\ee
Combining (\ref{4.11})--(\ref{4.13}), we obtain
\bea
\left(\frac{r^2 u_r^2}{\sqrt{1+fY}}\right)_r &=&\left(\frac{r^2}f\left(\sqrt{1+fY}-1\right)\right)_r+\frac{r u_r^2}{\sqrt{1+fY}}+\frac{N^2 u^2}{r\sqrt{1+fY}}\nn\\
&&-\frac{2r}f\left(\sqrt{1+fY}-1\right)+2r^2 V' uu_r.\label{4.14}
\eea
Using (\ref{4.10}), we have
\be\label{4.15}
\int_0^R \left(\frac{r^2}f\left(\sqrt{1+fY}-1\right)\right)_r\,\dd r=\frac{N^2 u_\infty^2}2 +\mbox{O}(R^{-1}).
\ee
Consequently, integrating (\ref{4.14}) over $0<r<R$, using the boundary condition on $u$, and applying (\ref{3.17}) and (\ref{4.15}), we arrive at
the quantization formula
\be\label{4.16}
\lim_{R\to\infty}\left\{2\int_0^R\left(\frac1f\left(\sqrt{1+fY}-1\right)+V\right) r\dd r-\int_0^R\frac{rY}{\sqrt{1+fY}}\,\dd r\right\}=\frac{N^2 u^2_\infty}2,
\ee
thus rendering an accurate refinement of the virial relation (\ref{4.9}), in the same spirit of the development in the previous two sections.

As a concrete example, we consider the following complexified AdS throat model \cite{ST} given by
\be\label{4.17}
f(|\phi|^2)=\alpha|\phi|^{-4},\quad \alpha>0.
\ee
With the Mexican-hat potential density (\ref{2.8}) to realize a spontaneously broken symmetry and the vortex ansatz
(\ref{2.11}), the energy density (\ref{4.4}) becomes the radially reduced one
\be\label{4.18}
{\cal E} (u)=\frac{u^4}\alpha\left(\sqrt{1+\frac\alpha{u^4}\left[u_r^2+\frac{N^2}{r^2}u^2\right]}-1\right)+\frac\lm 8(u^2-1)^2,
\ee
such that the equation of motion (\ref{4.8}) assumes the form
\bea\label{4.19}
&&\left(\frac{ru_r}{\sqrt{1+\frac\alpha{u^4}\left(u_r^2+\frac{N^2}{r^2} u^2\right)}}\right)_r=\frac{N^2 u}{r\sqrt{1+\frac\alpha{u^4}\left(u_r^2+\frac{N^2}{r^2} u^2\right)}}\nn\\
&&+\frac{4r u^3}{\alpha}\left(\sqrt{1+\frac\alpha{u^4}\left(u_r^2+\frac{N^2}{r^2} u^2\right)}-1\right)
-\frac{2r\left(u_r^2+\frac{N^2}{r^2}u^2\right)}{u\sqrt{1+\frac\alpha{u^4}\left(u_r^2+\frac{N^2}{r^2} u^2\right)}}+\frac{\lm r}2(u^2-1) u,\quad\quad
\eea
subject to the boundary condition (\ref{2.14}), and  the quantization rule (\ref{4.16}) reads
\bea\label{4.20}
&&\lim_{R\to\infty}\int_0^R\left\{\frac{2u^4}\alpha\left(\sqrt{1+\frac\alpha{u^4}\left(u_r^2+\frac{N^2}{r^2} u^2\right)}-1\right)+\frac\lm4(u^2-1)^2
-\frac{\left(u_r^2+\frac{N^2}{r^2} u^2\right)}{\sqrt{1+\frac\alpha{u^4}\left(u_r^2+\frac{N^2}{r^2} u^2\right)}}\right\}\,r\dd r\nn\\
&&=\frac{N^2}2,
\eea
 respectively. Computational results of this problem will also be presented in Section 6.

\section{Vortices in Dirac--Born--Infeld tachyon scalar field model}
\setcounter{equation}{0}

Interestingly, we may also extend our investigation to the complexified tachyon scalar field model governed by the Lagrangian action density \cite{CST,FKS,GNS,JK}
\be\label{5.1}
{\cal L}=- V(|\phi|^2)\sqrt{1-\pa_\mu\phi\pa^\mu\overline{\phi}}=V(|\phi|^2)\left(1-\sqrt{1-\pa_\mu\phi\pa^\mu\overline{\phi}}
\right)-{V(|\phi|^2)}.
\ee
The equation of motion may be found to be
\be
\frac12\,\pa_\mu\left(\frac{V(|\phi|^2)\pa^\mu\phi}{\sqrt{1-\pa_\nu\phi\pa^\nu\overline{\phi}}}\right)=-\sqrt{1-\pa_\nu\phi
\pa^\nu\overline{\phi}}\,\,V'(|\phi|^2)\phi.
\ee
Furthermore, we see that the (symmetrized) energy-momentum tensor is
\be
T_{\mu\nu}=\frac{V(|\phi|^2)(\pa_\mu\phi\pa_\nu\overline{\phi}+\pa_\mu\overline{\phi}\pa_\nu\phi)}{2\sqrt{1-\pa_\alpha\phi\pa^\alpha\overline{\phi}}}
-\eta_{\mu\nu}{\cal L}.
\ee
Thus, the energy density for a static field assumes the form
\be\label{5.4}
{\cal E}=V(|\phi|^2)\sqrt{1+|\nabla\phi|^2},
\ee
and the equation of motion becomes
\be\label{5.5}
\nabla\cdot\left(\frac{V(|\phi|^2)\nabla\phi}{\sqrt{1+|\nabla\phi|^2}}\right)=2\sqrt{1+|\nabla\phi|^2}\,V'(|\phi|^2)\phi.
\ee
Using the conservation law of the energy-momentum tensor, we obtain the energy-partition relation
\be\label{5.6}
\int_{\bfR^n}\frac{V(|\phi|^2)|\nabla\phi|^2}{\sqrt{1+|\nabla\phi|^2}}\,\dd^n x=n\int_{\bfR^n} V(|\phi|^2)\sqrt{1+|\nabla\phi|^2}\,\dd^n x.
\ee

We now specialize in two dimensions and consider the $N$-vortex solution given in (\ref{2.11}). Then (\ref{5.4}) and (\ref{5.6}) become
\bea
&&{\cal E}={\cal E}(u)=V(u^2)\sqrt{1+u_r^2+\frac{N^2}{r^2}u^2},\label{5.7}\\
&&\int_0^\infty \frac{V(u^2)\left(u_r^2+\frac{N^2}{r^2}u^2\right)}{\sqrt{1+u_r^2+\frac{N^2}{r^2}u^2}}\,r\dd r=2\int_0^\infty
V(u^2)\sqrt{1+u_r^2+\frac{N^2}{r^2}u^2}\,\, r\dd r,\label{5.8}
\eea
respectively, so that (\ref{5.5}) reduces to
\be\label{5.9}
\left(\frac{V(u^2) r u_r}{\sqrt{1+u_r^2+\frac{N^2}{r^2}u^2}}\right)_r= \frac{V(u^2)N^2 u}{r\sqrt{1+\frac{N^2}{r^2}u^2}}
+2r\sqrt{1+u_r^2+\frac{N^2}{r^2}u^2}\,\,V'(u^2) u.
\ee
Note also that
\be\label{5.10}
ru_r 
\left(\frac{Vr u_r}{\sqrt{1+u_r^2+\frac{N^2}{r^2} u^2}}\right)_r=
\left(\frac{V r^2 u_r^2}{\sqrt{1+u_r^2+\frac{N^2}{r^2} u^2}}\right)_r-
\frac{V ru_r}{\sqrt{1+u_r^2+\frac{N^2}{r^2} u^2}}(u_r+ru_{rr}),
\ee
and that
\bea\label{5.11}
\frac{VN^2 uu_r}{\sqrt{1+u_r^2+\frac{N^2}{r^2}u^2}}&=&\left(r^2 V\sqrt{1+u_r^2+\frac{N^2}{r^2} u^2}\right)_r-2r(V+rV'uu_r)
\sqrt{1+u_r^2+\frac{N^2}{r^2} u^2}
 \nn\\
&&+\frac{rV\left(\frac{N^2}{r^2} u^2 -ru_r u_{rr}\right)}{\sqrt{1+u_r^2+\frac{N^2}{r^2} u^2}}.
\eea
Thus, multiplying (\ref{5.9}) by $ru_r$ and applying (\ref{5.10}) and (\ref{5.11}), we obtain
\bea\label{5.12}
\left(\frac{r^2 V u_r^2}{\sqrt{1+u_r^2+\frac{N^2}{r^2}u^2}}\right)_r&=&\left( r^2 V {\sqrt{1+u_r^2+\frac{N^2}{r^2}u^2}}\right)_r
-2rV {\sqrt{1+u_r^2+\frac{N^2}{r^2}u^2}}\nn\\
&&+\frac {rV}{\sqrt{1+u_r^2+\frac{N^2}{r^2}u^2}}\left(u_r^2+\frac{N^2}{r^2}u^2\right),
\eea
which appears rather neat. Now, integrating (\ref{5.12}) over $(0,R)$ and letting $R\to\infty$, we find
\be\label{5.13}
\lim_{R\to\infty}\int_0^R\left(2V(u^2){\sqrt{1+u_r^2+\frac{N^2}{r^2}u^2}}-\frac{V(u^2)\left(u_r^2+\frac{N^2}{r^2}u^2\right)}{\sqrt{1+u_r^2+\frac{N^2}{r^2}u^2}}\right)\, r\dd r=\frac{N^2}2 V(u^2_\infty)u_\infty^2.
\ee

The relation (\ref{5.13}) implies the following alternatives.
\begin{enumerate}
\item[(i)] When $V(u^2_\infty)u^2_\infty>0$, we see that the state $u_\infty$ cannot serve as a ground state, an $N$-vortex solution
satisfying the boundary condition (\ref{3.15}) necessarily carries infinite energy, both
sides of (\ref{5.8}) are divergent, and (\ref{5.13}) refines (\ref{5.8}) with a square-of-the-vortex-number law. 
\item[(ii)] When $V(u^2_\infty)u^2_\infty=0$,
we see that the state $u_\infty$ is a ground state and an $N$-vortex solution satisfying the boundary condition (\ref{3.15}) may carry a finite energy so that
both sides of (\ref{5.8}) are finite which coincide with (\ref{5.13}) (with vanishing right-hand side).
\end{enumerate}

As a concrete example,
we may consider the rolling massive scalar model \cite{C,GNS} given by 
\be\label{5.15}
V(|\phi|^2)=V_0\e^{\pm\frac12 M^2|\phi|^2}.
\ee
In this case the model does not possess a finite ground state. Nevertheless, we may still {\em prescribe} an asymptotic state to acquire an $N$-vortex solution. 
In this situation, the equation of motion (\ref{5.9}) assumes the specific form
\be\label{5.16}
\left(\frac{ru_r}{\sqrt{1+u_r^2+\frac{N^2}{r^2}u^2}}\right)_r=\frac{r}{\sqrt{1+u_r^2+\frac{N^2}{r^2}u^2}}
\left(\frac{N^2}{r^2}\mp M^2 u_r^2\right)u\pm M^2 r {\sqrt{1+u_r^2+\frac{N^2}{r^2}u^2}}\,\, u,
\ee
and the conservation law (\ref{5.13}) becomes
\be\label{5.17}
\lim_{R\to\infty}\int_0^R\left(2\e^{\pm\frac12 M^2 u^2}{\sqrt{1+u_r^2+\frac{N^2}{r^2}u^2}}-\frac{\e^{\pm\frac12 M^2 u^2}\left(u_r^2+\frac{N^2}{r^2}u^2\right)}{\sqrt{1+u_r^2+\frac{N^2}{r^2}u^2}}\right)\, r\dd r=\frac{N^2}2 \e^{\pm\frac12 M^2 u_\infty^2}u_\infty^2,
\ee
both being independent of the ``initial level" of the potential density, $V_0$, at the vortex core.

\section{Computational examples}
\setcounter{equation}{0}

In this section, we present some numerical examples of the vortex solutions to the Born--Infeld wave equations studied in the previous sections. For our problems,
the equations are of the form $u''=f(r,u,u')$ ($r>0$), subject to the boundary condition $u(0)=0, u(\infty)=u_\infty$.
 With the notation $y=u,z=u'$, we may recast the two-point boundary value problem into
\be\label{6.1}
y'=z,\quad z'=f(r,y,z),\quad r>0;\quad y(0)=0,\quad y(\infty)=u_\infty,
\ee
where $f(r,y,z)$ is singular at $r=0$, which causes  difficulty. Another difficulty is that the interval over which we construct the solutions extends to infinity. These
difficulties will be overcome numerically where we work on an interval of the form $(r_0,r_1)$ for which $r_0>0$ is sufficiently small and $r_1>r_0$ sufficiently large
so that we may instead reformulate from (\ref{6.1}) a regular initial value problem:
\be\label{6.2}
y'=z,\quad z'=f(r,y,z),\quad r_0<r<r_1;\quad y(r_0)=a,\quad z(r_0)=b,
\ee
where $a,b$ are parameters. We shall choose $a,b$ suitably so that $y(r)$ stays close to zero for $r$ near $r_0$ and $y(r)$ approaches $u_\infty$ as $r$ tends to
$r_1$. Methodologically, we may use the classical fourth-order Runge--Kutta scheme to solve (\ref{6.2}). Besides, to facilitate our search for suitable $a,b$, we
employ the optimization functions stored in MATLAB to solve the problem
\be\label{6.3}
\min_{a,b} \left\{ R(a,b)=(y(r_1)-u_\infty)^2\right\}.
\ee
After a solution is obtained, the information below $r_0$ will be assessed using the asymptotic estimates of the solution there and that beyond $r_1$ will be
neglected. In fact, a more elaborated consideration on the asymptotic behavior of the solution $u$ enables us to employ the form $u(r)=u_0 r^N$ for small $r$
which  renders $a=u_0 r^N_0$ and $b=Nu_0 r_0^{N-1}$, as a consequence. In such a way, the two-parameter optimization problem (\ref{6.3}) may now be
simplified into the following one-parameter optimization problem
\be
\min_{u_0}\left\{R(u_0)=(y(r_1)-u_\infty)^2\right\},
\ee
which eases our computation.
 Finally, various energetic quantities are evaluated by numerical integration of the solutions over their respective intervals.

In our concrete examples below, we illustrate the results obtained for the equations (\ref{2.13}), (\ref{326}), (\ref{328}), and (\ref{4.19}). For all our purposes, we
may choose $r_0=0.0001$ and $r_1=30$.

\begin{figure}[h!]
   \centering
   \includegraphics[scale=0.6]{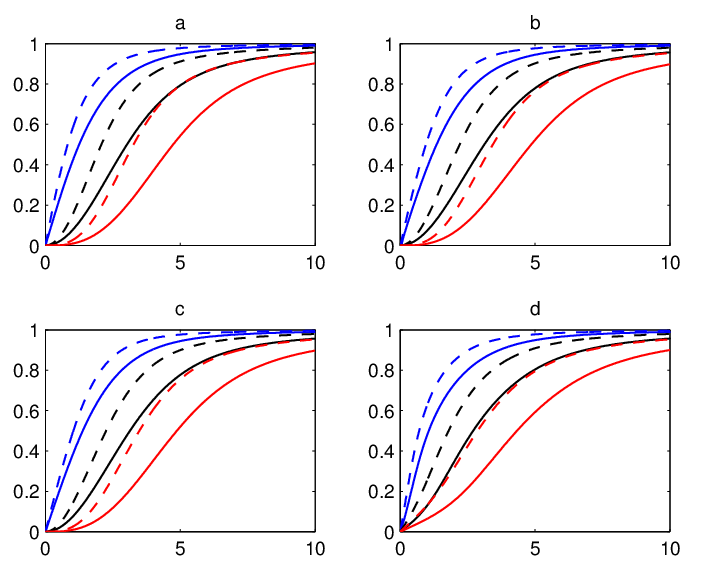}
   \caption{The plots of the solutions of the equations  (\ref{2.13}) ($b=1$), (\ref{326}) ($\Lambda=10$), (\ref{328}), and (\ref{4.19}) ($\alpha=1$), for
$\lm=1$ (solid curves) and $\lm=2$ (dashed curves), and $N=1,2,3$.}\label{F1}
\end{figure}

In Figure \ref{F1}, we present the computational results for solutions of (\ref{2.13}) (with $b=1$), (\ref{326}) (with $\Lambda=10$), (\ref{328}), and (\ref{4.19})
(with $\alpha=1$), plotted in a, b, c, d,
respectively. The solid curves are the solutions with $\lm=1$ and dashed ones with $\lm=2$, which are placed from upper to lower, with the winding number
$N=1,2,3$, grouped in a sequential order. Although computed from the Born--Infeld wave equations of quite different properties and difficulties, the solutions behave similarly.

We now consider the quantization identities (\ref{2.24a}), (\ref{3.23}), (\ref{3.25}), (\ref{4.20}) for the computed solutions.
In Table \ref{T1}, we compare the radial energy gaps, $\delta$, given on the left-hand sides of these identities, against the exact values of $\frac{N^2}2$, stated as
the right-hand sides of the identities. It is seen that these quantities match impressively well in all situations. In particular, we observe that, up to numerical errors, these
quantities are indeed insensitive to the fine structure of various models.


\begin{table}[h]
\centering 
\begin{tabular}{cccccccc} 
$\lm$ & $N$ & ${N^2}/2$&  $\delta$ in (\ref{2.24a})& $\delta$ in (\ref{3.23})&$\delta$ in (\ref{3.25})&$\delta$ in (\ref{4.20})\\ 
  \hline
      1 & 1 & 0.5 & 0.499893 &0.499914 &0.499792 &0.499327\\
      2 & 1 & 0.5 & 0.499682 &0.499875 &0.499848 &0.498954\\
      1 & 2 & 2   & 1.997384 &1.994599 &1.998325 &1.999168\\
      2 & 2 & 2   & 1.996918 &1.997356 &1.999214 &1.998753\\
      1 & 3 & 4.5 & 4.497325 &4.493439 &4.497132 &4.498306\\
      2 & 3 & 4.5 & 4.498817 &4.496771 &4.495863 &4.494813\\
\hline 
\end{tabular}
\caption{Comparison of the theoretical quantities given by the exact quantization laws and the numerical results based on
the computed solutions of the equations  (\ref{2.13}) ($b=1$), (\ref{326}) ($\Lambda=10$), (\ref{328}), and (\ref{4.19}) ($\alpha=1$), for
$\lm=1,2$ and $N=1,2,3$. }\label{T1}  
\end{table}

It will be instructive to know how the reduced radial Higgs potential energy 
\be\label{6.4}
P(u)=\int_0^\infty \frac\lm4 (u^2-1)^2\,r\dd r,
\ee
depends on the input data for our computed solutions, since it appears as a common quantity in  (\ref{2.24a}), (\ref{3.23}), (\ref{3.25}), and (\ref{4.20}),
and is conserved in the Ginzburg-Landau equation situation as stated in (\ref{GL}), giving rise to
\be\label{6.5}
P(u)=\frac{N^2}2.
\ee

Table \ref{T2} below lists our results. It is clear that, for fixed $N$, $P$ decreases
as $\lm$ increases in (\ref{2.24a}) and (\ref{4.20}), but increases in (\ref{3.23}) and (\ref{3.25}). In particular, unlike in (\ref{GL}) or (\ref{6.5}),
in the Born--Infeld situation $P$ depends on $\lm$ and other fine structures of the models
sensitively.

\begin{table}[h]
\centering 
\begin{tabular}{cccccccc} 
$\lm$ & $N$ & ${N^2}/2$&  $P$ in (\ref{2.24a})& $P$ in (\ref{3.23})&$P$ in (\ref{3.25})&$P$ in (\ref{4.20})\\ 
\hline 
      1 & 1 & 0.5 & 0.457306 &0.500385 &0.514944 &0.395297\\
      2 & 1 & 0.5 & 0.416537 &0.500818 &0.547693 &0.367217\\
      1 & 2 & 2   & 1.891836 &1.995613 &2.015481 &1.718639\\
      2 & 2 & 2   & 1.807552 &1.999396 &2.054653 &1.619042\\
      1 & 3 & 4.5 & 4.285314 &4.495453 &4.518372 &3.878193\\
      2 & 3 & 4.5 & 4.089426 &4.500814 &4.582015 &3.612965\\
\hline 
\end{tabular}
\caption{Numerical computations of the reduced radial version of the Higgs potential energy $P$ given in (\ref{6.4}) for 
the solutions of the equations  (\ref{2.13}) ($b=1$), (\ref{326}) ($\Lambda=10$), (\ref{328}), and (\ref{4.19}) ($\alpha=1$), for
$\lm=1,2$ and $N=1,2,3$, respectively, which appears as a common quantity on the left-hand sides of (\ref{2.24a}), (\ref{3.23}), (\ref{3.25}),
and (\ref{4.20}), respectively. }\label{T2}  
\end{table}

\section{Conclusions}
\setcounter{equation}{0}

In this paper, we have found a class of quantization identities valid for quite generic types of the Born--Infeld theories that possess vortices. The identities can intuitively be understood as a ``correction" to the virial theorem, or rather, that subtracting the two sides of the standard virial theorem renders a quantized ``discrepancy" quantity proportional to ${N^2}$, where $N$ is the winding number or vortex charge of the vortex solution, which would be zero in the usual,
finite-energy-mechanics, situations. In the case of global vortices in the Born--Infeld theories, the energy is infinite and the virial theorem is trivially valid as both sides diverge. We find the above-mentioned discrepancy quantity to be a universal quantized constant, proportional to ${N^2}$, present to provide a
refined energetic structure. This paper has shown a range of examples providing evidence for the fact that such a hidden but precise structure does not depend on the details of the theoretical formalism, but solely on the existence of the nonvanishing winding number in theories with a Born--Infeld kinetic term so that the energy is transversely divergent. The examples we have discussed include the Born--Infeld wave equations and related models considered in cosmology and 
finally the Born--Infeld tachyon models. The importance of the result lies in the universality of the theorem. It does not depend on the details of the potentials, the coupling parameters, 
and other structural properties of the models.

Summarizing, it may be enlightening and  worthwhile to recast the quantization identities derived in the previous sections in general terms. For convenience, use $V(|\phi|^2)$ to denote
a general potential density function with a spontaneous broken symmetry characterized by a normalized vacuum state, $|\phi|=1$. That is, $V\geq0, V(1)=0$.
Then it is clear that, without resorting to radial symmetry, for an $N$-vortex solution of the Born--Infeld wave equation (\ref{2.3}) over $\bfR^2$
the quantization law (\ref{2.24a}) assumes the generalized form
\be\label{7.1}
\lim_{R\to\infty}\int_{|x|<R}\left\{b^2\left(\sqrt{1+\frac1{b^2}|\nabla\phi|^2}-1\right)+V(|\phi|^2)-\frac{|\nabla\phi|^2}{2\sqrt{1+\frac1{b^2}|\nabla\phi|^2}}\right\}\,\dd^2 x=\frac{\pi N^2}2,
\ee
or even more generally, for that of the equation (\ref{3.6}), the corresponding identity (\ref{3.21}) reads
\be\label{7.2}
\lim_{R\to\infty}\int_{|x|<R}\left\{-F\left(-\frac12|\nabla\phi|^2\right)+V(|\phi|^2)-\frac12 F'\left(-\frac12|\nabla\phi|^2\right)|\nabla\phi|^2\right\}\,\dd^2 x=\frac{\pi N^2}2.
\ee
Another extended version of (\ref{7.1}) is (\ref{4.16}), arising from a Dirac--Born--Infeld model in the study of inflationary cosmology, for an $N$-vortex solution 
governed by the wave equation (\ref{4.5}), whose general form without imposing radial symmetry, is
\bea\label{7.3}
&&\lim_{R\to\infty}\int_{|x|<R}\left\{\left(\frac1{f(|\phi|^2)}\left[\sqrt{1+f(|\phi|^2)|\nabla\phi|^2}-1\right]+V(|\phi|^2)\right)-\frac{|\nabla\phi|^2}{2\sqrt{1+f(|\phi|^2)|\nabla\phi|^2}}\right\}\,\dd^2 x\nn\\
&&=\frac{\pi N^2}2,
\eea
where $f$ is a positive-valued function. All these are in the form of the law (\ref{1.1}). For the Dirac--Born--Infeld tachyon scalar field model (\ref{5.1}), an $N$-vortex
solution of (\ref{5.5}), approaching its asymptotic state $\eta$ at infinity ($|\phi(x)|\to\eta$ as $|x|\to\infty$),  satisfies (\ref{5.13}) under radial symmetry, or in general, the law
\be\label{7.4}
\lim_{R\to\infty}\int_{|x|<R}\left\{V(|\phi|^2)\sqrt{1+|\nabla\phi|^2}-\frac{V(|\phi|^2)|\nabla\phi|^2}{2\sqrt{1+|\nabla\phi|^2}}\right\}\,\dd^2 x=\frac{\pi N^2}2V(\eta^2)\eta^2,
\ee
which is also in the token of (\ref{1.1}). Although the left-hand sides of (\ref{7.1})--(\ref{7.4}) are complicated in view of the fine details contained in various models,
the right-hand sides of these identities are universally simple. In particular, they are independent of the detailed structures of the models.

Finally, among the many prospects of possible future studies along the line of the research reported in this work, it will be interesting and of value to unveil quantized energy 
conservation laws in higher-order powers of the vortex charge for various Born--Infeld type models considered here and elsewhere, and similarly quantized
laws in dimensions higher than two, as fruitfully developed in our earlier work \cite{GGY} for conventional theories.

\medskip

{\bf Acknowledgments.} ZG was partially supported by Natural Science Foundation of China under Grant No. U1504102.
SBG was supported by the Ministry of Education, Culture, Sports, Science (MEXT)-Supported Program for the Strategic Research Foundation at Private Universities ``Topological Science" (Grant No. S1511006) and by a Grant-in-Aid for Scientific Research on Innovative Areas ``Topological Materials Science" (KAKENHI Grant No. 15H05855) from MEXT, Japan.
YY was partially supported
by Natural Science Foundation of China under Grant No. 11471100.

\end{document}